\documentclass{aa}  

\usepackage{natbib,twoopt}
\usepackage{multirow}
\usepackage{amsmath}
\usepackage{amssymb} 
\usepackage[T1]{fontenc}
\usepackage{blindtext}
\usepackage{hyperref}
\hypersetup{
    colorlinks=true,
    linkcolor=red,
    filecolor=blue,
    citecolor=blue,
    urlcolor=blue,
    pdftitle={Overleaf Example},
    pdfpagemode=FullScreen,
}
\usepackage{graphicx}
\usepackage{gensymb}
\usepackage{subcaption}
\usepackage{pythonhighlight}
\usepackage{listings}
\usepackage{xcolor}
\usepackage{multirow}
\usepackage{siunitx}
\usepackage{makecell}
\usepackage{lscape}
\usepackage{orcidlink}

\usepackage{amsmath} 
\newcommand{\angstrom}{\text{\normalfont\AA}}

\usepackage[varg]{txfonts}

\usepackage{graphicx}	
\usepackage{amsmath}	
\usepackage{tablefootnote}
\usepackage{mathtools}

\definecolor{codegreen}{rgb}{0,0.6,0}
\definecolor{codegray}{rgb}{0.5,0.5,0.5}
\definecolor{codepurple}{rgb}{0.58,0,0.82}
\definecolor{backcolour}{rgb}{0.95,0.95,0.92}
\defcitealias{MRN1977}{MRN}
\defcitealias{wd2001}{WD01}
\defcitealias{hm2020}{HM20}

\lstdefinestyle{mystyle}{
  backgroundcolor=\color{backcolour}, commentstyle=\color{codegreen},
  keywordstyle=\color{magenta},
  stringstyle=\color{codepurple},
  basicstyle=\ttfamily\footnotesize,
  breakatwhitespace=false,         
  breaklines=true,                 
  captionpos=b,                    
  keepspaces=true,                                               
  showspaces=false,                
  showstringspaces=false,
  showtabs=false,                  
  tabsize=2
}

\begin{document}

\title{Investigating interstellar dust along the line of sight of GX 13+1 using different dust size distributions}

\author{B. Vaia \orcidlink{0000-0003-0852-0257}\inst{1,2,3,4},
       S. T. Zeegers\orcidlink{0000-0002-8163-8852}\inst{4,5,8},
       I. Abril-Cabezas \orcidlink{0000-0003-3230-4589}\inst{6,7},
       E. Costantini\orcidlink{0000-0001-8470-749X}\inst{4,8}}
          
\authorrunning{B. Vaia et al.}

\institute{Scuola Universitaria Superiore IUSS Pavia, Piazza della Vittoria 15, 27100 Pavia, Italy\\ \email{beatrice.vaia@inaf.it}
  \and Department of Physics, University of Trento, Via Sommarive 14, 38123 Povo (TN), Italy 
  \and Istituto Nazionale di Astrofisica, Istituto di Astrofisica Spaziale e Fisica Cosmica di Milano, via A. Corti 12, 20133 Milano, Italy
  \and SRON Space Research Organisation Netherlands, Niels Bohrweg 4, 2333 CA Leiden, The Netherlands
  \and European Space Agency (ESA), European Research and Technology Centre (ESTEC), Keplerlaan 1, 2201 AZ Noordwijk, The Netherlands
  \and DAMTP, Centre for Mathematical Sciences, University of Cambridge, Wilberforce Road, Cambridge CB3 0WA, United Kingdom
  \and Kavli Institute for Cosmology Cambridge, Madingley Road, Cambridge, CB3 0HA, United Kingdom
  \and Anton Pannekoek Institute for Astronomy, University of Amsterdam, Science Park 904, NL-1098 XH Amsterdam, the Netherlands
}
   \date{Received: 22/10/2025; accepted: 27/04/2026}


\abstract
{High-resolution X-ray spectroscopy offers a powerful tool to investigate the physical and chemical properties of dust grains, especially through the analysis of absorption edges of elements such as oxygen, magnesium, silicon, and iron, which are the main constituents of interstellar dust. In all previous X-ray studies, these absorption edges have been modeled assuming the MRN grain size distribution. This model successfully reproduces the average interstellar extinction curve. However, with the advent of new observations, it shows important limitations, indicating that more complex grain-size distributions are required to fully describe interstellar dust properties.}
{We aim to constrain the composition and size distribution of interstellar dust along the line of sight to the bright low-mass X-ray binary GX 13+1.}
{We analyzed high-resolution X-ray spectra obtained with the \textit{Chandra} HETG instrument (MEG+1 and MEG$-$3) and simultaneously modeled the Si K and Mg K absorption edges. For the first time, we compared the classical Mathis et al. 1977, ApJ, 217, 425 grain size distribution with other grain size distributions, thus exploring different ISM densities.}
{Our analysis rules out scenarios of both very diffuse and very dense ISM, favoring grain size distributions associated with average Galactic conditions along this line of sight. The dust composition is found to be dominated by amorphous olivine and the crystallinity contribution is about 2\%. The depletion patterns and elemental abundances derived are consistent with prior X-ray and infrared studies.}
{}


\maketitle
\nolinenumbers
\section{Introduction}\label{sec1}

Understanding the composition and properties of interstellar dust remains a fundamental challenge in astrophysics. Interstellar dust is characterized by the composition, morphology, and size distribution of its various particles, and by the abundance, relative to hydrogen, of its elemental constituents. 

Over time, models of interstellar dust have progressively evolved in response to increasingly detailed observational data. One of the foundational models, proposed by \citet{MRN1977} (hereafter \citetalias{MRN1977}), successfully reproduced the average interstellar extinction curve. This model describes dust as bare, spherical silicate and graphite grains distributed in size according to a power law with radii ($a$) ranging from $0.005\,\rm{\mu m}$ to $0.25\,\rm{\mu m}$.
However, observations from the \textit{Infrared Astronomical Satellite} (IRAS) all-sky survey revealed limitations in the \citetalias{MRN1977} model, particularly an unexpected excess of $12\,\rm{\mu m}$ and $25\,\rm{\mu m}$ emission from the diffuse interstellar medium (ISM), indicating that the \citetalias{MRN1977} distribution alone could not fully account for the observed dust emission. This discrepancy led to the introduction of polycyclic aromatic hydrocarbon (PAH) molecules as an additional dust component, as first proposed by \citet{ATB1985}. Observations from the \textit{Cosmic Background Explorer} (COBE) confirmed the need for small carbonaceous grains, which were subsequently included in dust models by \citet{wd2001} (hereafter \citetalias{wd2001}) and further investigated in more recent works such as \citet{hm2020} 
(hereafter \citetalias{hm2020}).

Regarding the composition, dust is generally categorized into carbonaceous and silicate grains, with additional components such as oxides (e.g., MgO, $\rm{SiO_2}$), carbides (e.g., SiC), and metallic iron \citep{Draine2011}. Infrared spectral analysis of silicate grains indicates a preference for Mg-rich compositions with stoichiometries between those of olivine and pyroxene \citep{Kemper2004, Min2007, Fogerty2016}. Furthermore, infrared observations suggest that less than 2\% of interstellar dust in the Galaxy exists in crystalline form \citep{Kemper2004,DoDuy2020}. On the other hand, the Stardust Interstellar Dust Collector \citep{Westphal2014} showed a large fraction of crystalline material, probably preserved inside larger particles. 

Dust has a strong impact on many astronomical observations. First, it causes the extinction of radiation, mainly in the optical and UV wavebands, which, for extragalactic objects, requires a correction accounting for dust both in our Galaxy and in the host galaxy. Second, the thermal emission of interstellar dust contaminates astronomical observations at long wavelengths, significantly affecting, in particular, the study of the Cosmic Microwave Background (CMB).
Dust also plays a critical role in X-ray astronomy, as it absorbs and scatters high-energy photons. 
The advent of modern X-ray observatories, such as XMM-\textit{Newton} and \textit{Chandra}, has enabled detailed spectroscopic studies of these interactions (e.g. \citealt{Lee2002,Costantini2012,Pinto2013,Zeegers2019,Rogantini2019,Rogantini2020}). 
The presence of solid particles modifies the X-ray absorption edges of abundant metals, producing oscillatory modulations known as X-ray Absorption Fine Structure (XAFS, \citealt{Costantini2022}). These features are sensitive to both the composition and the crystallinity of the dust grains, and even to the dust composition. This makes XAFS a powerful and independent probe of the solid phase of the interstellar medium, complementary to extinction, and infrared-based studies. Models of the XAFS are based on laboratory measurements, in particular the Si K edge has been studied by \citet{Zeegers2017} performing Synchrotron measurements as well as the Mg K edge studied by \citet{Rogantini2019}.

The spectra of X-ray binaries provide an excellent opportunity to study the absorption of dust and gas along their line of sight.
Galactic Low Mass X-ray Binaries (LMXBs) within our Galaxy are particularly well-suited for investigating XAFS due to their their high flux, significant column densities, and relatively simple spectra, with most sources showing no intrinsic absorption features below 10 keV.

In this paper, we study the Si K and Mg K absorption edges (1.84 keV and 1.3 keV) observed with the \textit{Chandra} High-Energy Transmission Grating Spectrometer (HETG) toward the bright LMXB GX 13+1. For the first time, we directly compare the traditional \citetalias{MRN1977} distribution with more complex grain size models, including the \citetalias{wd2001} distribution for three different values of the ratio of total to selective extinction, $R_V$, (3.1, 4.0 and 5.5) and the \citet{hm2020} distribution for four distinct dense gas fractions, $\eta_{\rm{dense}}$ (0.1, 0.3, 0.5, and 0.8). In addition, we exploit for the first time the superior spectral resolution of the third-order HETG spectrum at the Si K edge, enabling a more robust characterization of the absorption profile and of the dust properties along this line of sight. We further note that this is the first analysis not affected by the instrumental feature present in all other spectra (see Appendix \ref{appA}).

The paper is organized as follows: Section \ref{dustdistribution} introduces the dust size distribution model, Section \ref{sec3} outlines the data reduction process, Section \ref{sec4} presents the analysis methodology and results, Section \ref{sec5} discusses the implications of our findings, and Section \ref{sec6} summarizes our conclusions.

\section{Dust size distributions models}
\label{dustdistribution}

\begin{table}
    \centering
    \caption{Dust samples used in this paper, with their chemical formula, form, and references.}
    \label{table:dustsample}
    \begin{tabular}{lccc}
    \hline
    Name & Chemical formula & Form & References \\
    \hline\hline
    Olivine & $\rm{MgFeSiO_4}$ & amorphous & [2],[3] \\
    Olivine & $\rm{Mg_{1.56}Fe_{0.4}Si_{0.91}O_4}$ & crystalline & [1],[2],[3] \\
    Fayalite & $\rm{Fe_2SiO_4}$ & crystalline & [2] \\
    Forsterite & $\rm{Mg_2SiO_4}$ & crystalline & [2],[3] \\
    Quartz & $\rm{SiO_2}$ & crystalline & [3] \\
    Quartz & $\rm{SiO_2}$ & amorphous & [3] \\
    Quartz & $\rm{SiO_2}$ & amorphous & [3] \\
    Spinel & $\rm{MgAl_2O_4}$ & crystalline & [3],[4] \\
    Enstatite & $\rm{MgSiO_3}$ & amorphous &  [2],[3] \\
    Enstatite & $\rm{MgSiO_3}$ & crystalline &  [2],[3] \\
    En75Fe25 & $\rm{Mg_{0.75}Fe_{0.25}SiO_3}$ & amorphous & [2],[3]  \\
    En60Fe40 & $\rm{Mg_{0.6}Fe_{0.4}SiO_3}$ & amorphous & [1],[2],[3] \\
    En60Fe40 & $\rm{Mg_{0.6}Fe_{0.4}SiO_3}$ & crystalline & [1],[2],[3] \\
    En90Fe10 & $\rm{Mg_{0.9}Fe_{0.1}SiO_3}$ & amorphous & [1],[2],[3] \\
    En90Fe10 & $\rm{Mg_{0.9}Fe_{0.1}SiO_3}$ & crystalline & [1],[2],[3] \\
    \hline
    \end{tabular}
    \tablefoot{[1] \citet{Rogantini2018} [2] \citet{Zeegers2019}; [3] \citet{Rogantini2020}; [4] \citet{Costantini2019}.
    The amorphous quartz samples are distinguished into intermediate and fully amorphous structures.}
\end{table}

The X-ray silicate spectra adopted in this work are based on laboratory measurements presented in \citet{Zeegers2017,Zeegers2019} and \citet{Rogantini2020}. Table~\ref{table:dustsample} shows the dust species used in this study, together with their chemical composition and structural state (crystalline or amorphous).
The adaptation of the laboratory measurements to X-ray spectra follows the general procedure described in \citet{Zeegers2017} and \citet{Rogantini2018}. Here we summarize the procedure and for details we refer to the respective studies. 

\begin{itemize}
    \item From the laboratory measurements, the optical constants were derived for each dust species~\citep{Zeegers2017,Zeegers2019,Rogantini2020}. The dimensionless optical constants describe how light propagates through a specific material at each wavelength.  
    \item The anomalous diffraction theory ~\citep[ADT,][]{vdHulst1957} is employed to obtain the absorption $Q_\mathrm{abs}(a,\lambda)$, scattering $Q_\mathrm{sca}(a,\lambda)$ and extinction efficiencies $Q_\mathrm{ext}(a,\lambda)$, at each wavelength ($\lambda$) and grain size ($a$). 
    \item The extinction cross section at each grain size and wavelength is defined as $C_\mathrm{ext}(a,\lambda)=\pi a^2Q_\mathrm{ext}(a,\lambda)$. The total extinction cross section $\sigma_\mathrm{ext}(\lambda)$ is obtained by integrating over a grain size distribution:
\begin{equation}
    \sigma_\mathrm{ext}=\int{C_\mathrm{ext}(a,\lambda)n_{\mathrm{gr}}da}
\end{equation}
In this equation $n_{\mathrm{gr}}$ is the dust grain size distribution. 
\item Finally, in order to compare the laboratory measurements with astronomical data, the resulting extinction cross section profiles are implemented in the \texttt{amol} model of the \texttt{SPEX} fitting code~\citep{Kaastra1996}.

\end{itemize}

In previous X-ray spectroscopy studies  of interstellar dust ~\citep[e.g.,][]{Zeegers2017, Rogantini2018}, the size distribution ($n_{\mathrm{gr}}$) of the dust grains  were calculated using an \citetalias{MRN1977} size distribution.

In this work, we compare the \citetalias{MRN1977} distribution with more complex dust size distribution models, such as those by \citetalias{wd2001} and \citetalias{hm2020}.

\begin{figure}
    \centering
    \includegraphics[width=0.49\textwidth]{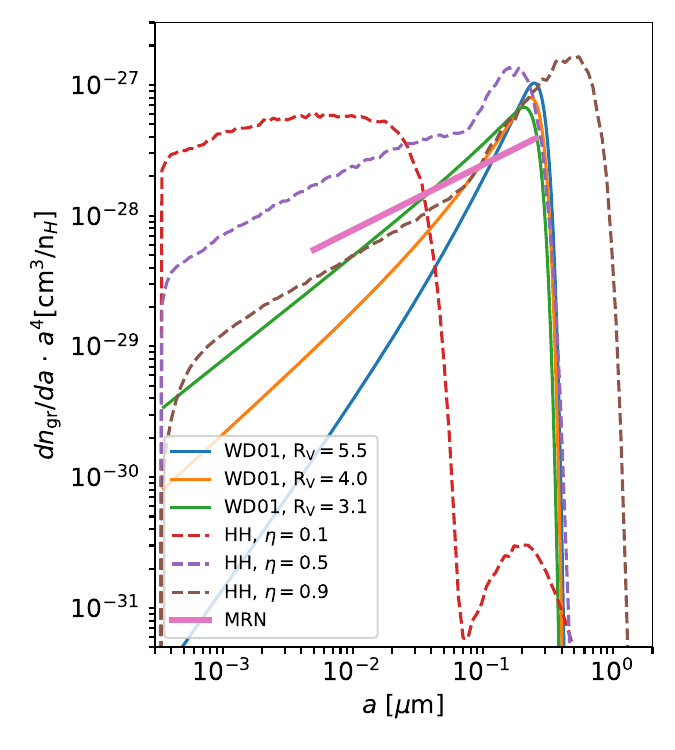}
    \caption{Comparisons of several different silicate size distributions used in this study. The thick pink line shows the MRN distribution. The blue, orange and green lines show the \citetalias{wd2001} models with $R_V$ of 5.5, 4.0 and 3.1. Three size distribution models of \citetalias{hm2020} are shown by the dashed red purple and brown lines, indicating different values of dense gas fraction with a dust evolution time of 10 Gyr.}
    \label{fig:sizedistrib}
\end{figure}

\begin{figure*}[h]
    \centering
    \begin{subfigure}[b]{0.45\textwidth}
        \centering
        \includegraphics[width=\textwidth]{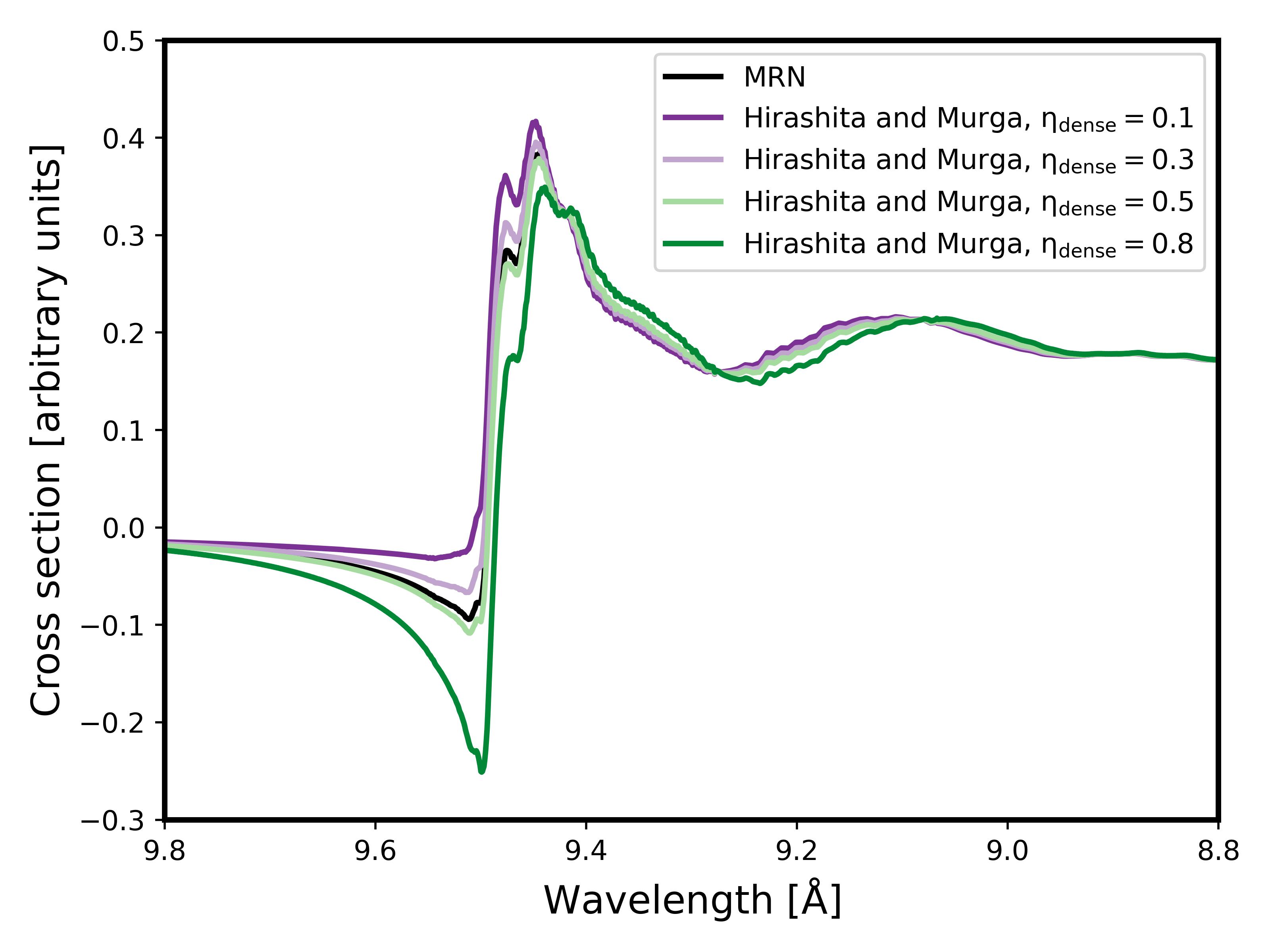}
        \label{fig:hm_mg}
    \end{subfigure}
    \hfill
    \begin{subfigure}[b]{0.45\textwidth}
        \centering
        \includegraphics[width=\textwidth]{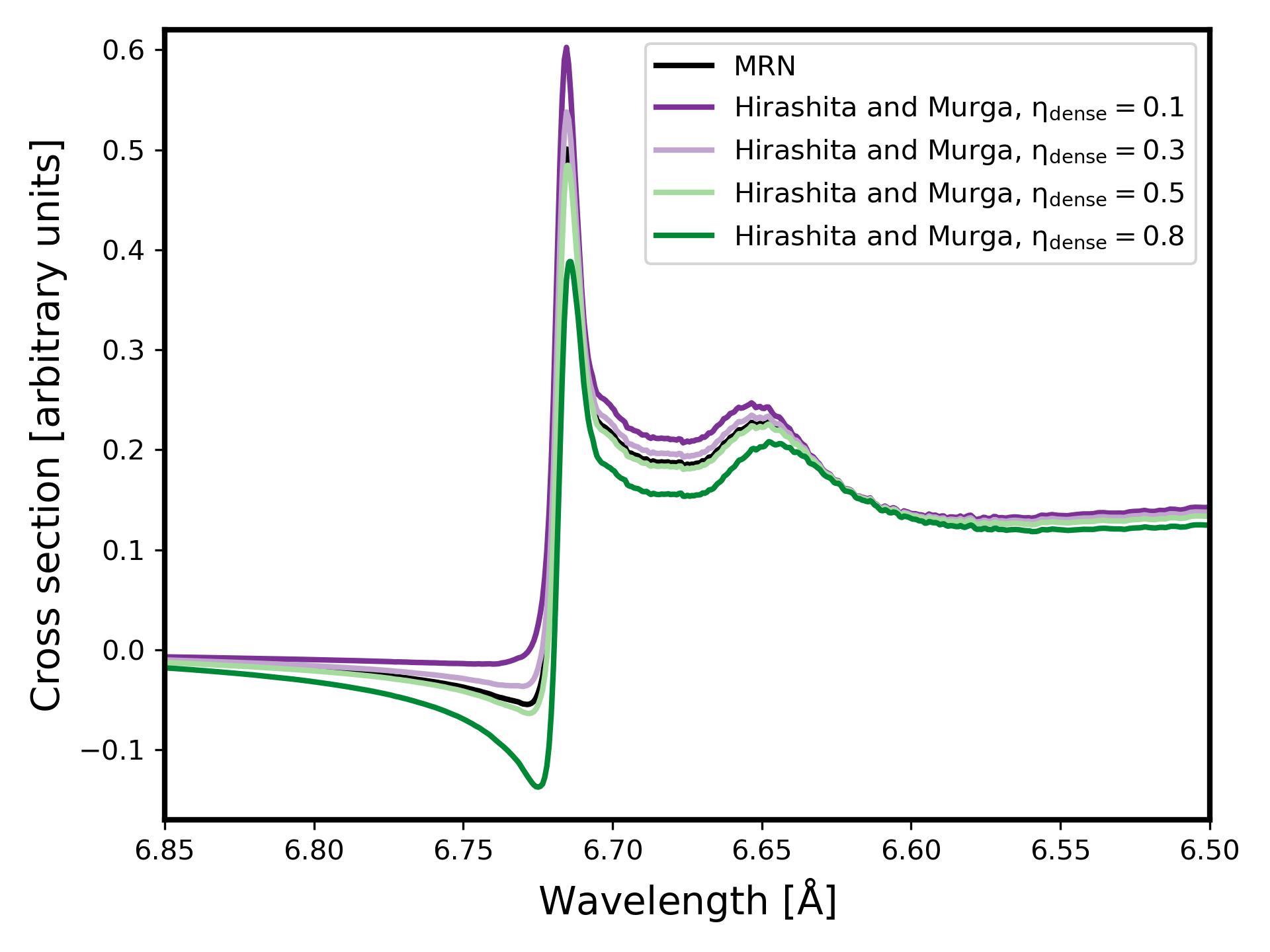}
        \label{fig:hm_si}
    \end{subfigure}

    \vspace{0.5cm}

    \begin{subfigure}[b]{0.45\textwidth}
        \centering
        \includegraphics[width=\textwidth]{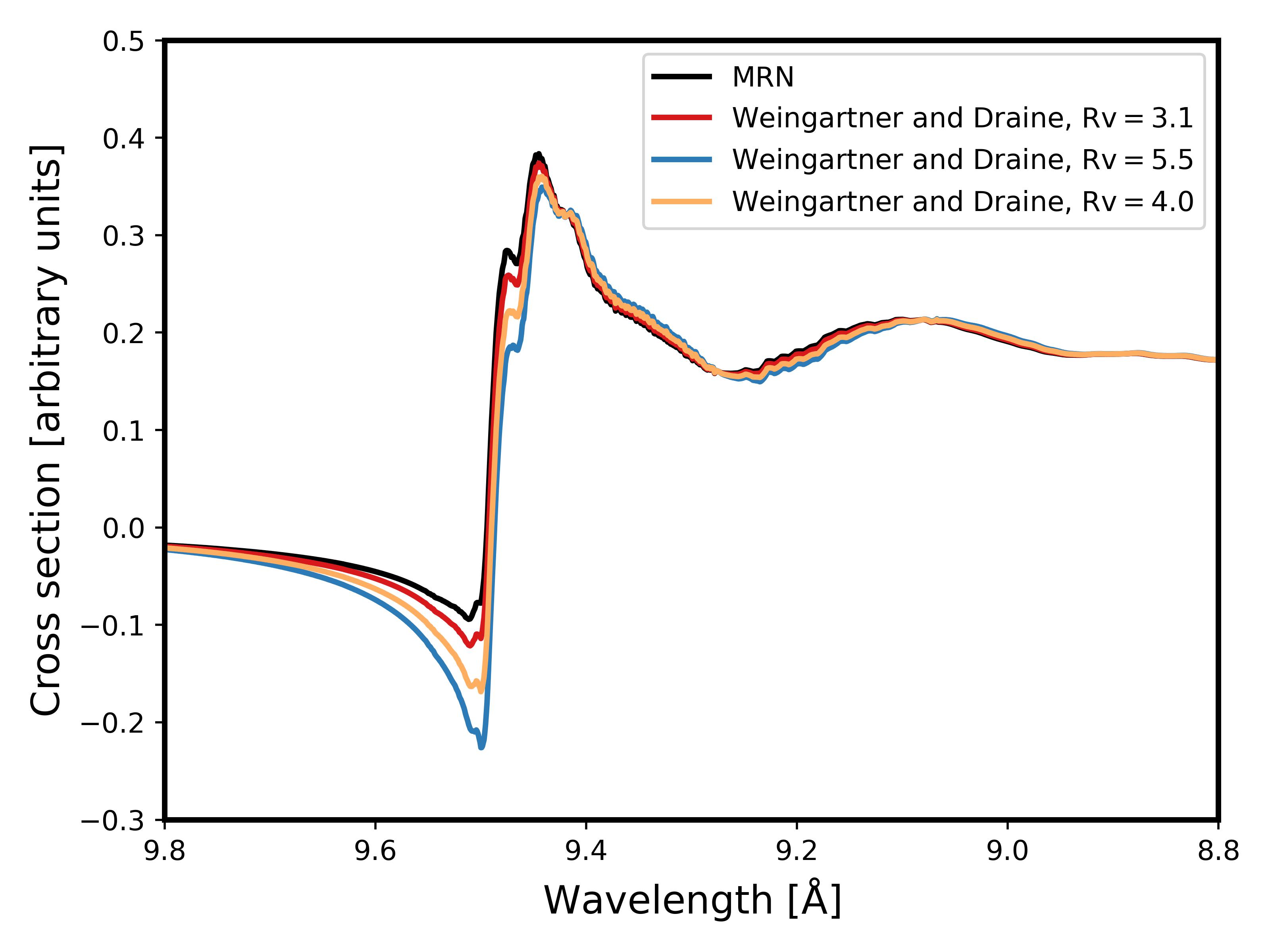}
        \label{fig:wd_mg}
    \end{subfigure}
    \hfill
    \begin{subfigure}[b]{0.45\textwidth}
        \centering
        \includegraphics[width=\textwidth]{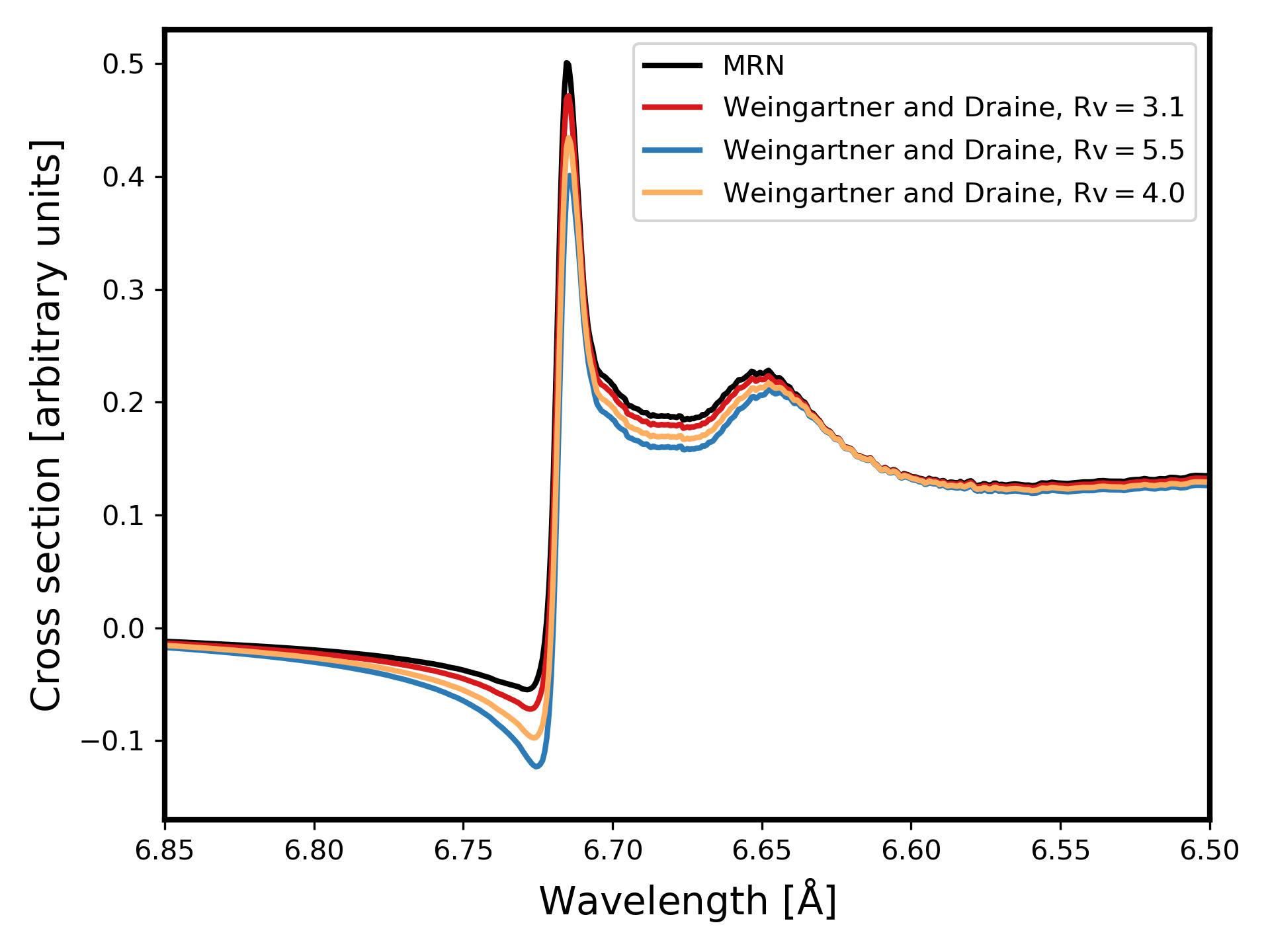}
        \label{fig:wd_si}
    \end{subfigure}

    \caption{Extinction cross sections of amorphous olivine calculated with different dust size distributions, shown around the magnesium edge (left column) and the silicon edge (right column). The top row compares the \citetalias{MRN1977} distribution (black) with that of \citetalias{hm2020} for dense gas fractions of 0.1 (purple), 0.5 (light purple and light green), and 0.8 (green). The bottom row compares the \citetalias{MRN1977} distribution (black) with that of \citetalias{wd2001} for reddening values of $R_V=3.1$ (red), $R_V=4.0$ (orange), and $R_V=5.5$ (blue).}
    \label{fig:MRNvsHM_WD}
\end{figure*}

These models do not account for mixed carbonaceous and silicate grains; rather, they adopt separate grain size distributions for the carbonaceous and silicate dust components. Examples of the size distribution models are shown in Fig.~\ref{fig:sizedistrib}. 
The corresponding extinction spectra at the energies of the magnesium ($\sim$1305 eV; 9.5 $\angstrom$) and silicon ($\sim$1839 eV; 6.74 $\angstrom$) absorption edges are presented in Fig.~\ref{fig:MRNvsHM_WD}. In this figure, we show several extinction cross sections for amorphous olivine.
These cross sections will be included in a future release of the \texttt{amol} model in \texttt{SPEX}~\citep{Kaastra1996}. 
In the following, we describe each of the size distribution models.

The simplest model is the \citetalias{MRN1977} model, shown by the pink line in Fig.~\ref{fig:sizedistrib}. This model was developed to describe the observed extinction by dust. \citetalias{MRN1977} found that the functional dependence of the extinction on the wavelength (i.e. the extinction curve) is well described by the following grain size distribution: 
\begin{equation}
dn_{\mathrm{gr}}(a) = C_s n_{\mathrm{H}}a^{-3.5}
da,\, a_{\mathrm{min}} < a < a_{\mathrm{max}} 
\end{equation}
Here the minimum grain size  $a_{min}=0.005\,\mu\mathrm{m}$ and the maximum grain size is $a_{max}=0.25\,\mu\mathrm{m}$, $C_s$ is a normalization constant, $n_{\mathrm{gr}}$ is the number density of grains and $n_{\mathrm{H}}$
is the number density of H nuclei; $dn_{\mathrm{gr}}$ can also be expressed as $n(a)da$. 
We follow the procedure described in~\citet{Rogantini2018} to calculate the normalization constant $C_s$, which is based on observations of X-ray halos from~\citet{Mauche1986}. \citet{Mauche1986} measured the excess surface brightness of X-ray sources at 1 keV, arising from dust scattering and therefore correlating with the dust column density. They derived the following relation (their Eq.~15):
\begin{equation}
n_{\mathrm{gr}}\sigma_{\mathrm{sca}}\approx 0.18\,\mathrm{kpc}^{-1}\quad \mathrm{at}\,1\,\mathrm{keV},
\end{equation}
where $n_{\mathrm{gr}}$ is the grain density and $\sigma_{\mathrm{sca}}$ the scattering cross section. 
 
To normalise the size distribution, \citet{Rogantini2018} adopt the correlation from \citet{Mauche1986}, computing the scattering cross section $\sigma_\mathrm{sca}$ at 1 keV for each dust species using ADT.

Although the \citetalias{MRN1977} model is a practical and simple parametrization of the size distribution, it does not always fit the observed extinction well. The extinction has been found to vary in the Galaxy and can be characterized by the ratio of total to selective extinction parameter $R_V$ ~\citep{CCM89}, which is defined as $R_V = A_V /E(B-V)$. This parameter correlates with the density of the ISM and the grain size: a lower $R_V$ corresponds to the diffuse ISM and indicates a larger population of small grains.

\citetalias{wd2001} models expand the size distribution to smaller particles than the classic \citetalias{MRN1977} distribution with a lower dust size limit of  $a_\mathrm{min} = 3.5\times 10^{-4}\,\mu\mathrm{m}$. The silicate grain size distribution is given by a function (equation 4 and 5 in \citetalias{wd2001}): 
\begin{equation}
    \begin{multlined}
    \frac{1}{n_{\mathrm{H}}}\frac{dn_{\mathrm{gr}}}{da} = \frac{C_s}{a}\left(\frac{a}{a_{t,s}}\right)^{\alpha_s}F(a;\beta_{s}a_{t,s})\\
    \times\begin{cases}
    1,\quad 3.5\times 10^{-4}\,\mu\mathrm{m} <a <a_{t,s}\\
    \mathrm{exp}\{-[(a-a_{t,s})/a_{c,s}]^3\},\quad a>a_{t,s}
    \end{cases}
    \end{multlined}.
\end{equation}
The silicate dust models contain a cut off ($a_{c,s}$) at $0.1\,\mu\mathrm{m}$, after which a turn off point of the curvature ($a_{t,s}$) of the grain sizes is introduced beyond $~0.1\,\mu\mathrm{m}$.
The term 
\begin{equation}
F(a;\beta_{s}a_{t,s})\equiv \begin{cases}
1+\beta a/a_{t,s}\quad \mathrm{if}\quad\beta\geq 0 \\
(1-\beta a/a_{t,s})^{-1}\quad \mathrm{if}\quad \beta< 0 
\end{cases},
\end{equation}
provides the curvature ($\beta$) to the distribution. The parameter $\alpha_s$ varies depending on the value of $R_V$ (among other parameters), between $\sim-1 \,\mathrm{and}\,-2$. 
We investigate these models using three different values of $R_V$ (3.1, 4.0 and 5.5). 
\citetalias{wd2001} normalised the size distribution models by fitting them to Galactic extinction curves, adopting the parametrization of~\citet{Fitzpatrick1999}, which depends on wavelength and $R_V$. The normalization of the curves, in turn, depends on the Galactic environment. For the diffuse ISM, where $R_V=3.1$, the normalization is provided by $A_V/N_\mathrm{H} = 5.3\times10^{-22}\,\mathrm{cm}^2$ (\citealt{Bohlin1978} and also applied in~\citealt{DraineLee1984}). 
In dense regions, where $R_V= 4.0$ or $5.5$, the extinction curve is normalized by $A_I/N_\mathrm{H} = 2.6\times10^{-22}\,\mathrm{cm}^2$ ($I=I$-band).  In addition, \citetalias{wd2001} assume an average "astronomical" silicate composition. We follow this parametrization for the silicates to normalize the size distribution.

The \citetalias{hm2020} model describes the evolution of the grain size distribution over 0.1, 0.3, 1.0, 3.0, and 10 Gyr. In our study we only considered the models with a dust evolution time of 10 Gyr, which resemble Milky Way dust. The silicate component in the \citetalias{hm2020} model accounts for about 60--70\% of the total dust mass. The minimum grain size considered is $a_\mathrm{min} = 3\times 10^{-4}\,\mu\mathrm{m}$. Particles smaller than this size are too small to be regarded as dust grains.
The largest grains used in their modeling have a radius $a_{\mathrm{max}}=10\,\mu\mathrm{m}$. The differences in gas density affects the grain growth and destruction mechanisms in the \citetalias{hm2020} dust evolution models. In this work we investigate the size distribution models for four different values of dense gas fraction ($\eta_{\rm{dense}}\,=\,$0.1, 0.3, 0.5, 0.8), exploring different environments with different densities.  
In \citetalias{hm2020} the interstellar medium is described as a two-phase medium composed of a diffuse (warm) component and a dense (cold) component, characterized by typical densities and temperatures of ($n_{\rm{H}}$, $T_{\rm{gas}}$) = (0.3 $\rm{cm}^{-3}, 10^4$ K) and (300 $\rm{cm}^{-3}$, 25 K), respectively. The dense gas fraction, $\eta_{\rm{dense}}$, therefore represents the fraction of the total ISM mass that resides in the dense, cold phase, as opposed to the diffuse warm phase. The normalization of the size distribution depends on the total dust mass. \citetalias{hm2020} assumes a closed box model. The galaxy starts with a gas mass with zero metallicity and converts the gas to stars with the total baryonic
(gas + stars) mass conserved. From the evolution models a total grain mass is obtained, which in turn is used to normalize the size distribution.
Similar to the models of \citetalias{wd2001}, smaller grains are present in the most diffuse environments, as can be seen in Fig.~\ref{fig:sizedistrib}.

\section{Observations and data reduction}\label{sec3}
\subsection{GX 13+1}

GX 13+1 is a luminous neutron star LMXB, originally classified as an atoll source due to its lower luminosity \citep{Hasinger1989}. However, subsequent observations have revealed properties more characteristic of Z sources \citep{Fridriksson2015}.
All \textit{Chandra} observations of GX 13+1 exhibit wind signatures \citep{Rogantini2025}, characterized by significantly blue-shifted absorption lines produced by an ionized outflow, indicating the presence of disk winds with velocities reaching up to approximately $800\,\rm{km\,s^{-1}}$ \citep{Ueda2004,Madej2014}.
GX 13+1 has previously shown type-I X-ray bursts \citep{Matsuba1995} and absorption dips \citep{Iaria2014}, but we did not observe any such events in our light curves.

Located at a distance of approximately 7 kpc from Earth \citep{Bandyopadhyay1999}, GX~13+1 lies at Galactic coordinates ($b = 0.11\degree$, $l = 13.52\degree$). Its high brightness ($\simeq 6 \times 10^{-9}\,\mathrm{erg\,cm^{-2}\,s^{-1}}$, \citealt{DAi2014}) enables the acquisition of high signal-to-noise spectra at the energies of the Mg and Si absorption edges, respectively at 1.30 and 1.84 keV. Moreover, the source lies in the Galactic plane, implying a large intervening column of interstellar material, which makes it a particularly well-suited target for probing interstellar dust through X-ray absorption features. Indeed, this source has already been included in samples aimed at studying interstellar dust absorption edges, especially those associated with silicon \citep{Zeegers2019,Yang2022} and sulfur \citep{Gatuzz2024}. Furthermore, GX~13+1 has been observed multiple times over the years, resulting in a large cumulative exposure time, making it one of the best targets for high-quality spectroscopic studies.

\subsection{Data reduction}
In this work we report the analysis of X-ray data collected by the HETG instrument of \textit{Chandra} \citep{Canizares2005} in TE mode between July 2010 and May 2019 (Obs.ID in Table \ref{table:obsid}). 

For each observation, we extracted the MEG +1 and MEG$-$3 spectra, and combined spectra of the same order across different observations using the CIAO tool \textsf{combine\_grating\_spectra} (version 4.15; \citealt{Fruscione2006}).
The data were reprocessed from level 1 event file, following the standard CIAO procedures. The spectra were extracted using \textsf{tgextract}, and the corresponding response matrices (RMFs) and ancillary response files (ARFs) were generated with \textsf{mkgrmf} and \textsf{fullgarf}.

These two orders (MEG+1 and MEG$-$3) are the only ones in which the silicon edge ($6.74\,\angstrom$) falls on a back-illuminated chip, thus avoiding the instrumental emission line reported by \citet{Rogantini2020} (see Appendix \ref{appA}).

The spectral analysis was restricted to the wavelength ranges $3\,\text{--}8\,\angstrom$ for the MEG$-$3 and $3\text{--}10\,\angstrom$ for the MEG+1. After this selection we inspected the spectra for pile-up. Pile-up arises when multiple photons are detected as a single event and is therefore frequently observed in the spectra of bright sources, such as GX 13+1.
The MEG+1 spectrum is affected by pile-up, particularly in the harder portion of the spectrum.\footnote{The magnesium and silicon edge regions are less susceptible to pile-up because ISM absorption significantly suppresses the spectrum in these areas.} Therefore, we excluded the MEG +1 data in the ranges $3.4\text{--}3.7\,\angstrom$ and $4.2\text{--}5.9\,\angstrom$, as these intervals exhibit systematic deviation from the continuum model.
Additionally, to retain the superior resolution of the MEG$-$3 spectrum, we also excluded the MEG+1 data around the silicon edge ($6.65\text{--}6.77\,\angstrom$).

\begin{table}[t]
	\centering
	\caption{Log of the \textit{Chandra} observations used in this paper.}
	\label{table:obsid}
	\begin{tabular}{ccc} 
    \hline
    Obs. ID & Start time (UTC) & Exposure (ks)\\
		\hline
        \hline
    11814 & 2010-08-01 T00:31:31 &  $29.1$\\
    11815 & 2010-07-24 T05:46:27 &  $29.0$\\
    11816 & 2010-07-30 T14:47:25 &  $29.1$\\
    11817 & 2010-08-03 T10:12:10 &  $29.1$\\
    20191 & 2018-05-10 T06:21:06 &  $25.1$\\
    20192 & 2018-05-26 T07:09:48 &  $23.1$\\
    20193 & 2019-02-17 T11:54:29 &  $26.2$\\
    20194 & 2019-05-30 T03:44:31 &  $25.4$\\
    \hline
	\end{tabular}
\end{table}

\section{Data analysis and results}\label{sec4}
We performed the spectral analysis using the code \texttt{spex} version 3.08 \citep{Kaastra1996}. All fits in this paper were performed using C-statistics \citep{Cash1979,Kaastra2017},  which can be used regardless of the number of counts per bin.
All errors are quoted at the 90\% confidence level for the parameter of interest.

\subsection{Continuum and absorption}

Following \citet{Zeegers2019}, we modeled the soft X-ray emission with a disk blackbody component (\texttt{dbb}; \citealt{Shakura1973}) and the hard X-ray emission with a Comptonization model (\texttt{comt}; \citealt{Titarchuk1994}). In the Comptonization model we fixed the temperature of the seed photons to that of the disk blackbody and left free to vary the  plasma temperature ($T_{\rm{comt}}$). Absorption by cold interstellar gas was accounted for using the \texttt{hot} model \citep{dePlaa2004}, with the temperature fixed at $1\times10^{-6}\,\rm{keV}$. We adopted the protosolar abundances from \citet{Lodders2010}, which are the default in \texttt{spex}.
To model the ionized outflowing gas from the source we added a \texttt{xabs} model \citep{Steenbrugge2003}, leaving free to vary the average systematic velocity of the absorber (\texttt{zv}) and the root mean square velocity (\texttt{v}). We adopted the default \texttt{xabs} input file provided by \texttt{spex} for the photoionisation balance parameters, which is based on photoionization calculations performed with Cloudy assuming the spectral energy distribution of NGC 5548 \citep{Steenbrugge2005}. This file provides pre-computed ionic column densities and electron temperatures as a function of the ionization parameter. Although previous works report multiple wind components in GX 13+1 \citep{Allen2018,Rogantini2025}, we verified that including a second \texttt{xabs} component does not improve the fit nor affect the Mg and Si absorption edges. A single photoionized absorber is therefore sufficient for the purposes of this work.
We simultaneously fit the continuum of the two datasets - one containing MEG+1 and the other MEG$-$3 data. To account for the different instrumental response, only the absorption parameters were linked between the two spectra.
We found a photo-ionized gas in outflow with a velocity $\texttt{zv}= (-4.7\,\pm\,1.8 )\times\;10^{2}\,\rm{km\,s^{-1}}$, $\texttt{v}=(3.0\,\pm\,0.5)\times10^{2} \,\rm{km\,s^{-1}}$, a logarithmic ionization parameter $\log \xi = 4.3_{-0.3}^{+0.1}$, and a hydrogen column density of $N_{\rm H} = (1.3\,\pm\,0.6)\times10^{23}\;\rm{cm^{-2}}$. These values are consistent with the one found by \citet{Zeegers2019}.
For the cold absorber, instead, we found a hydrogen column density of $N_{\rm H} = (3.4_{-0.6}^{+0.5})\times 10^{22}\;\rm{cm^{-2}}$, consistent with the value found in previous work \citep{Zeegers2019,Pintore2014,DAi2014}.
All the fitting parameters are reported in Table \ref{table:continuum_bestfit}. 

\begin{table}[ht]
\renewcommand{\arraystretch}{1.5}
\centering
\caption{Best-fit parameters derived for the continuum in the two different orders. The reported 2-10 keV flux is the observed (i.e., absorbed) flux, evaluated from the best-fitting model within the fitted energy range.}
\begin{tabular}{lcc}
\hline
\hline
 & MEG -3 & MEG +1 \\
\hline
$N_{\rm{H}}^{\rm{cold}}\,(10^{22}\,\rm{cm^{-2}})$ & \multicolumn{2}{c}{$3.4_{-0.6}^{+0.5}$}   \\
$k_{\rm{B}}T_{\rm{dbb}}\,(\rm{keV})$  &  \multicolumn{2}{c}{$0.7\,\pm\,0.1$}\\
$k_{\rm{B}}T_{\rm{comt}}\,(\rm{keV})$ & $18_{-9}^{+12}$ & $20\pm 15$ \\
$\tau_{\rm{comt}}\,(\rm{keV})$ & $0.9_{-0.1}^{+1.0}$ & $0.7_{-0.1}^{+1.4}$\\
$N_{\rm{H}}^{\rm{xabs}}\,(10^{23}\,\rm{cm^{-2}})$  &  \multicolumn{2}{c}{$1.3\,\pm\,0.6$}\\
$\log \xi^{\rm{xabs}}$  &  \multicolumn{2}{c}{$4.3_{-0.3}^{+0.1}$} \\
$\texttt{zv}_{\rm{out}}^{\rm{xabs}}\,(10^{2}\,\rm{km\,s^{-1}})$  &  \multicolumn{2}{c}{$-\,4.7\,\pm\,1.8$}  \\
$\texttt{v}^{\rm{xabs}}\,(10^{2}\,\rm{km\,s^{-1}})$  &   \multicolumn{2}{c}{$3.0\,\pm\,0.5$} \\
$F_{2-10\,\rm{keV}}\,(10^{-9}\,\rm{erg\,cm^{-2}\,s^{-1}})$  & \multicolumn{2}{c}{$8.5\,\pm\,0.4$} \\
\hline
\end{tabular}
\label{table:continuum_bestfit}
\end{table}

\subsection{The dust model}\label{da:dust}

After modelling the continuum of our source, along with the neutral and ionized absorption components, we included the dust model in the fit to simultaneously account for the magnesium and silicon absorption edges. As noticed by \citet{Rogantini2019}, fitting both edges simultaneously helps to better constrain the dust properties, reducing the potential degeneracies.
The \texttt{spex} \texttt{amol} model calculates the transmission of various molecules considering both absorption and scattering. By default the extinction models in \texttt{spex} are evaluated for grains with a standard \citetalias{MRN1977} size distribution. In this work we will test the grain size distribution models described in Section \ref{dustdistribution}.

The \texttt{amol} model supports the simultaneous fitting of mixtures comprising up to four distinct dust components. According to the approach outlined in \citet{Costantini2012}, the total number of possible compound combinations is given by:

\begin{equation}
C_{n,k} = \frac{n!}{k!(n-k)!}
\end{equation}

where $n$ represents the total number of available compounds (16 in this case, as listed in Table \ref{table:dustsample}) and $k$ is the number of components in each combination (4).
To fit the data, $C_{n,k}$ models are employed, with each unique combination representing a distinct extinction model. This process is repeated for each dust size distribution.

Following the approach in \citet{Rogantini2019,Rogantini2020}, we selected models that are statistically comparable to the best fit using the \textit{Akaike Information Criterion} (AIC, \citealt{Akaike1974}). According to the criteria outlined by \citet{Burnham2002}, models with a $\Delta\rm{AIC}<4$ are considered statistically indistinguishable from the best fit, while those with a $\Delta\rm{AIC}>10$ are excluded from further consideration.

In the subsequent section, we present the results obtained for each dust size distribution. The dust and gas column densities derived using the best-fit dust compositions for all distributions are summarized in Table \ref{table:dustgasNh}.
Note that the oxygen and iron gas column densities reported in Table \ref{table:dustgasNh} are derived indirectly, since the absorption edges of these elements fall outside the fitted energy range; consequently, the associated uncertainties are less robust than those of parameters directly constrained by the data.
\begin{table*}[]
\caption{Column densities of dust and gas derived from the best-fit models for different dust size distributions.}
\small
\renewcommand{\arraystretch}{1.3}
\resizebox{\textwidth}{!}{
\begin{tabular}{lcccccccc}
\hline
\hline
 & MRN & WD ($R_V = 3.1$)& WD ($R_V = 4.0$) & WD ($R_V = 5.5$) & HM ($\eta_{\rm{d}} = 0.1$) & HM ($\eta_{\rm{d}} = 0.3$) & HM ($\eta_{\rm{d}} = 0.5$) & HM ($\eta_{\rm{d}} = 0.8$) \\
\hline
$N_{\rm{a-olivine}}\,(10^{18}\,\mathrm{cm^{-2}})$ 
& $1.2^{+0.2}_{-0.5}$ & $0.9\,\pm\,0.2$ & $0.8\,\pm\,0.3$ & $0.5 \pm 0.1$ 
& $1.1^{+0.1}_{-0.2}$ & $1.0^{+0.1}_{-0.2}$ & $0.8^{+0.4}_{-0.5}$ & $0.5 \pm 0.1$ \\

$N_{\rm{a-quartz}}\,(10^{18}\,\mathrm{cm^{-2}})$ 
& $0.3^{+0.3}_{-0.1}$ & $0.4^{+0.2}_{-0.1}$ & $0.5\,\pm\,0.1$ & $0.9 \pm 0.1$ 
& $0.2^{+0.3}_{-0.1}$ & $0.3^{+0.3}_{-0.1}$ & $0.5 \pm 0.2$ & $0.9 \pm 0.1$ \\

$N_{\rm{c-forsterite}}\,(10^{18}\,\mathrm{cm^{-2}})$ 
& -- & -- & -- & -- & -- & -- & $0.1^{+0.2}_{-0.1}$ & -- \\
\hline
$N^{\rm{gas}}_{\rm{O}}\,(10^{19}\,\mathrm{cm^{-2}})$ 
& $1.2^{+0.6}_{-0.5}$ & $<1.8$ & $1.7^{+0.1}_{-0.7}$ & $<1.8$ 
& $1.3 \pm 0.2$ & $<1.8$ & $1.2^{+0.6}_{-0.3}$ & $1.2 \pm 0.3$ \\

$N^{\rm{gas}}_{\rm{Mg}}\,(10^{17}\,\mathrm{cm^{-2}})$ 
& $1.0^{+1.1}_{-1.0}$ & $2.2\pm1.2$ & $3.0 \pm 1.8$ & $2.4^{+2.6}_{-0.1}$ 
& $0.024 \pm 0.001$ & $1.2 \pm 0.1$ & $1.4^{+2.0}_{-1.4}$ & $1.3^{+3.5}_{-0.7}$ \\

$N^{\rm{gas}}_{\rm{Si}}\,(10^{17}\,\mathrm{cm^{-2}})$ 
& $0.2\pm0.1$ & $0.3\pm0.1$ & $0.4\pm0.2$ & $0.4 \pm 0.2$ 
& $0.1 \pm 0.1$ & $0.2^{+0.2}_{-0.1}$ & $0.2 \pm 0.1$ & $0.5 \pm 0.2$ \\

$N^{\rm{gas}}_{\rm{Fe}}\,(10^{16}\,\mathrm{cm^{-2}})$ 
& $0.2^{+1.9}_{-0.1}$ & $<2.1$ & $<2.1$ & $<2.1$ 
& $<2.1$ & $<2.1$ & $<2.1$ & $<2.1$ \\
\hline
C-stat & 1677 & 1681 & 1696 & 1708 & 1692 & 1677 & 1681 & 1719 \\
dof    & 1464 & 1464 & 1464 & 1464 & 1464 & 1464 & 1464 & 1464 \\
\hline
\end{tabular}
}

\tablefoot{
Column densities of dust and gas derived from the best-fit spectral models. 
In all cases, as described in Sect.~\ref{da:dust}, four dust components are fitted simultaneously. 
Except for the \citetalias{hm2020} model with $\eta_{\rm d}=0.5$, only amorphous olivine and amorphous quartz contribute significantly to the fit.
}

\label{table:dustgasNh}
\end{table*}

\normalsize

\paragraph{MRN distribustion.}

\begin{figure*}[h]
    \centering
    \begin{subfigure}[b]{0.45\textwidth}
        \includegraphics[width=\textwidth]{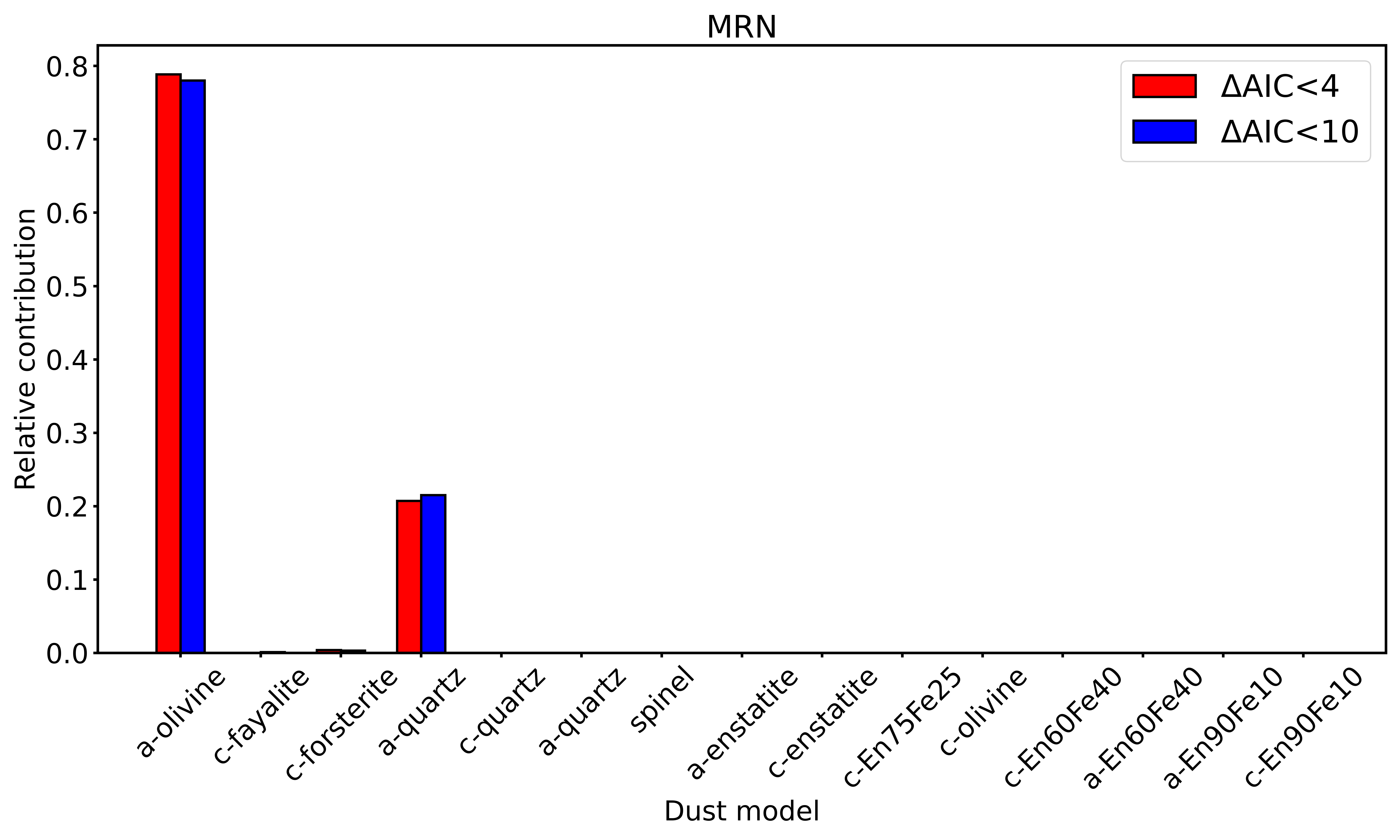}
    \end{subfigure}
    \hfill
    \begin{subfigure}[b]{0.45\textwidth}
        \includegraphics[width=\textwidth]{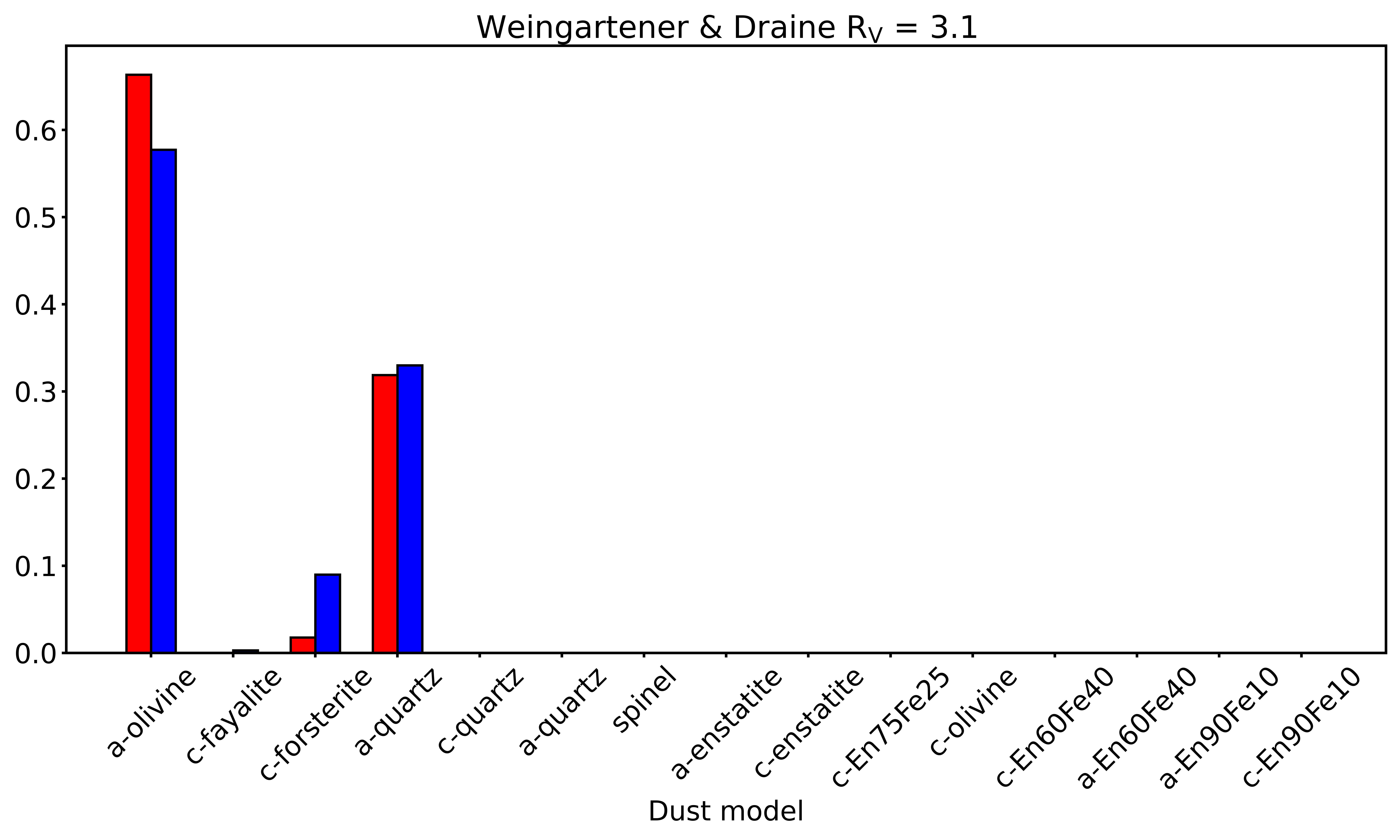}
    \end{subfigure}

    \begin{subfigure}[b]{0.45\textwidth}
        \includegraphics[width=\textwidth]{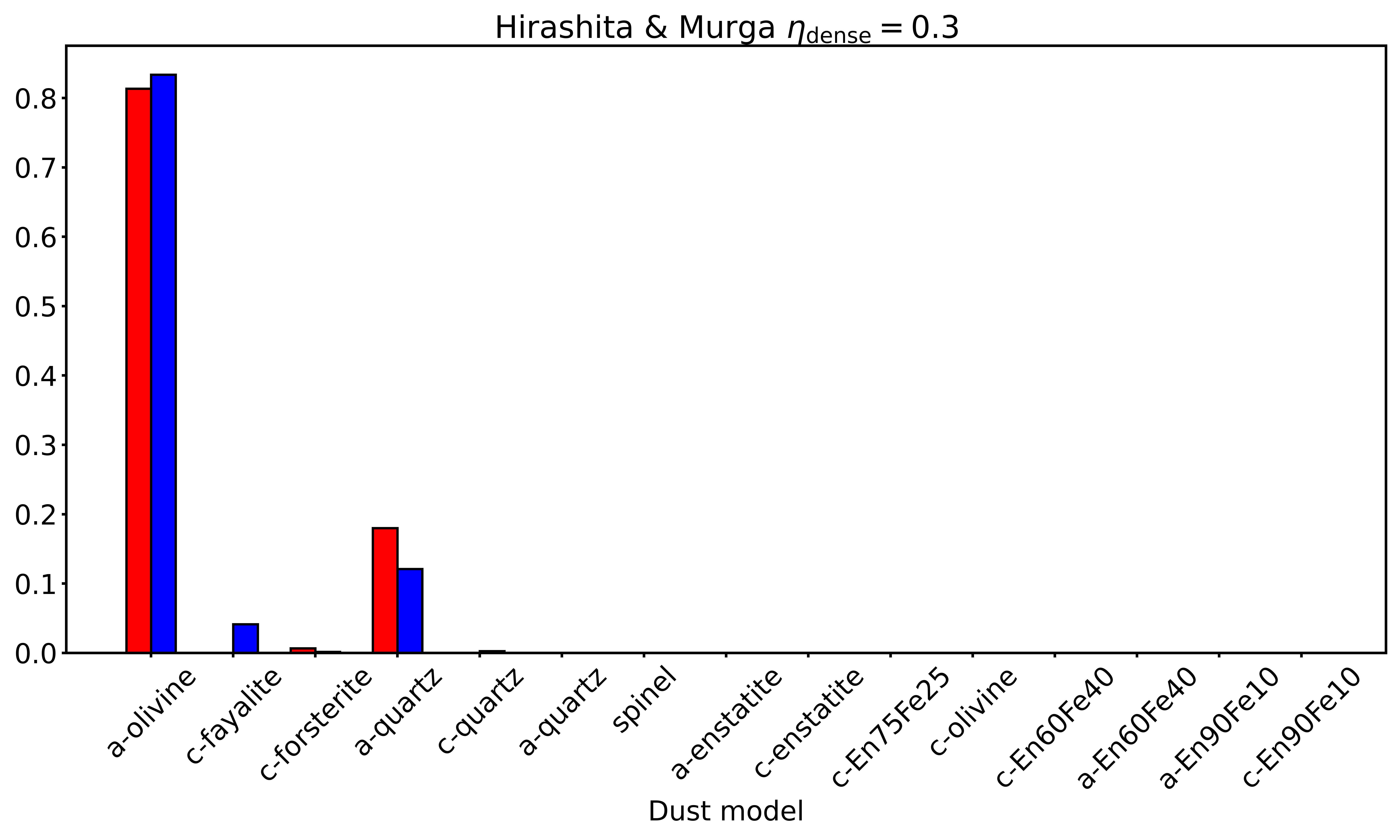}
    \end{subfigure}
    \hfill
    \begin{subfigure}[b]{0.45\textwidth}
        \includegraphics[width=\textwidth]{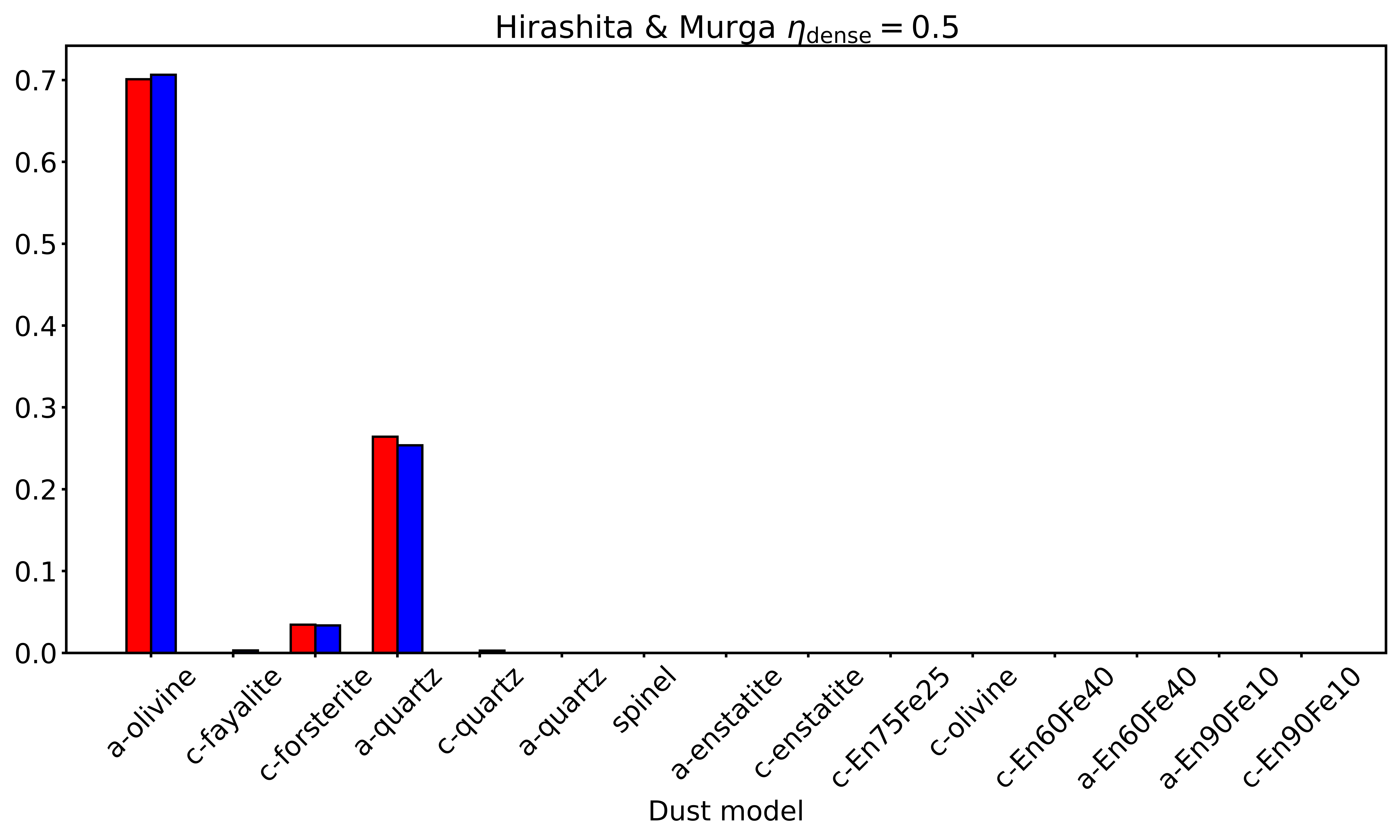}
    \end{subfigure}

    \caption{Histograms of the relative contribution of each dust species, based on AIC-selected models. Each panel shows results computed using different dust size distributions for the extinction cross section calculations.}
    \label{risultati_totali}
\end{figure*}

The initial analysis employed the standard \citetalias{MRN1977} size distribution for the extinction cross section, used to model the edge profiles (black curves in Fig. \ref{fig:MRNvsHM_WD}). The dust composition providing the closest match to the observed data primarily comprised amorphous olivine ($\sim 80\%$), with a smaller contribution from fully amorphous quartz $\sim 20\%$, see the top left panel in Fig.\ref{risultati_totali}. The model yielding the best statistical fit resulted in a C-statistic of 1677 with 1464 degrees of freedom.

\begin{figure*}[h]
    \centering
    \begin{subfigure}[b]{0.49\textwidth}
        \includegraphics[width=\textwidth]{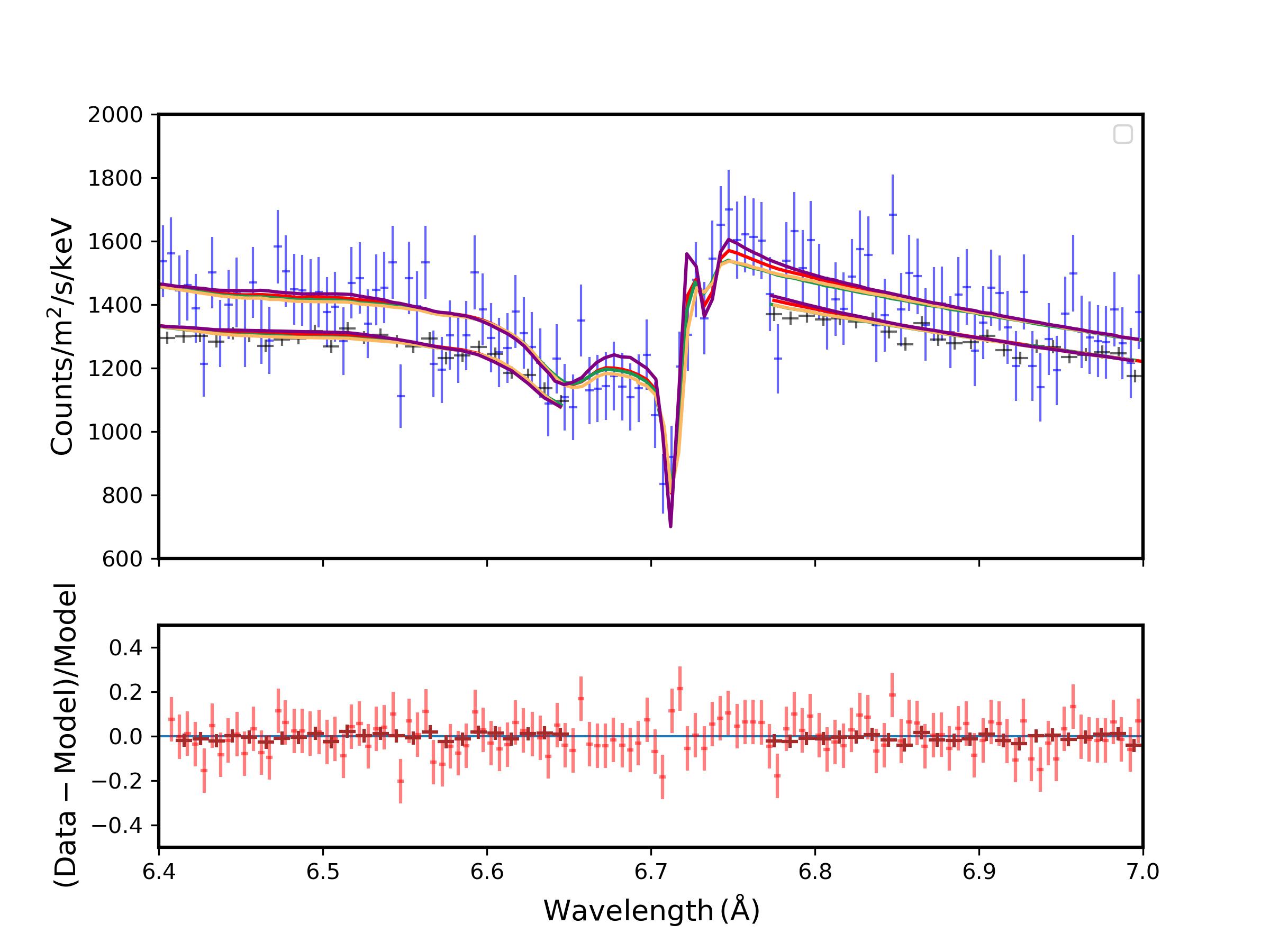}
        \label{fig:Si}
    \end{subfigure}
    \begin{subfigure}[b]{0.49\textwidth}
        \includegraphics[width=\textwidth]{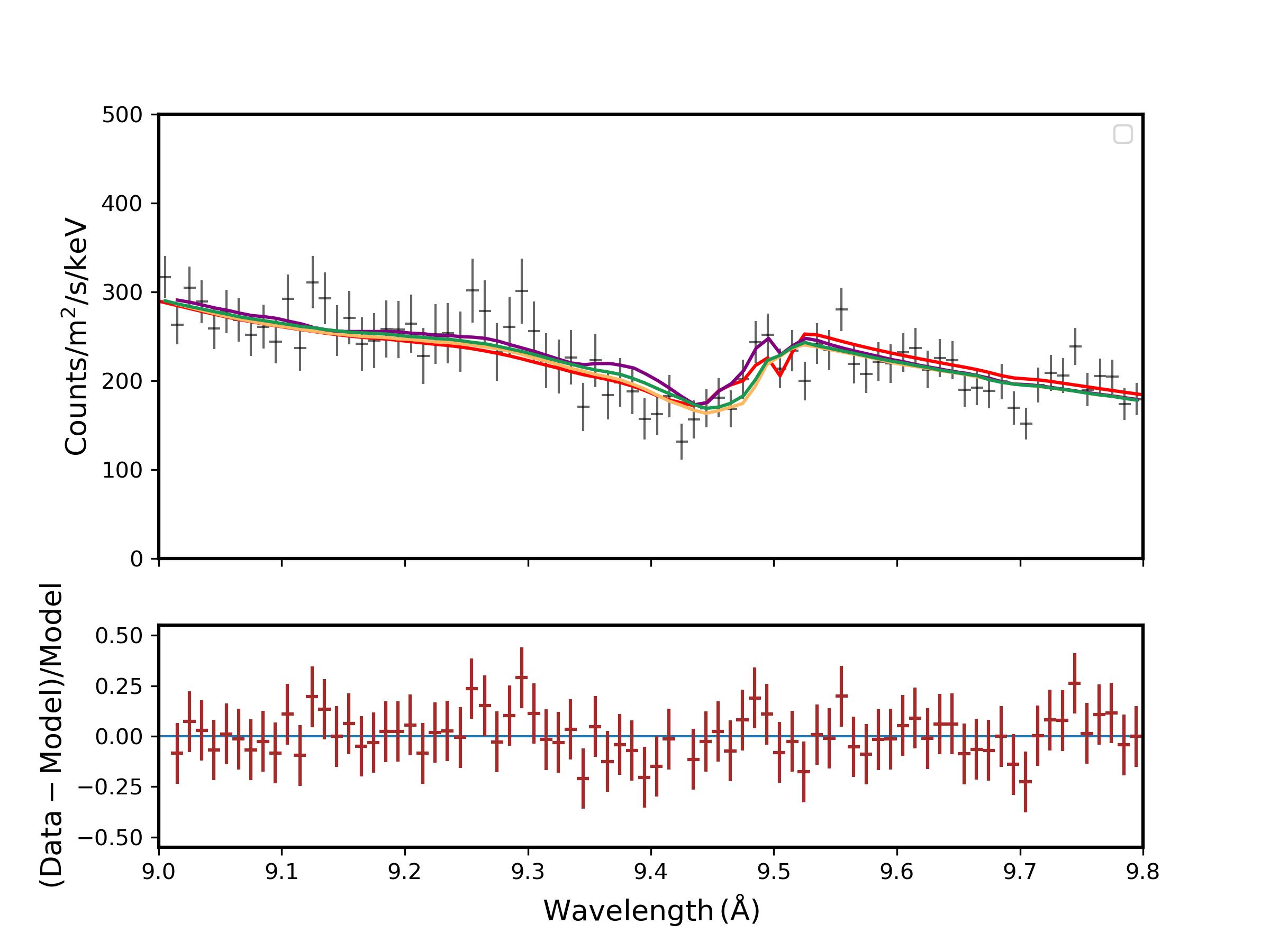}
        \label{fig:Mg}
    \end{subfigure}
    \caption{Top panel: Best  fits of the Si K edge (left panel) and Mg K edge (right panel) using the dust size distribution from \citetalias{wd2001} with $R_V = 3.1$ (in red), \citetalias{MRN1977} (in yellow), \citetalias{hm2020} with $\eta_{\rm{dense}} = 0.3$ (in green), and \citetalias{hm2020} with $\eta_{\rm{dense}} = 0.8$ (in purple). \textit{Chandra} MEG-3 data are shown in blue, and MEG+1 data in black. Bottom panel: Residuals defined as (observed-model)/model of the best fit obtained using \citetalias{MRN1977} dust size distribution.}
    \label{fig:WDbestfit_edges}
\end{figure*}

\paragraph{Weingartner \& Draine.}

The application of the dust size distribution from \citetalias{wd2001} with $R_V = 3.1$ (red curves in Fig. \ref{fig:MRNvsHM_WD}) resulted in a best-fit composition predominantly consisting of amorphous olivine ($\sim$ 70\%), with a complementary component of fully amorphous quartz ($\sim$ 30\%), as illustrated in the Top right panel in Fig. \ref{risultati_totali}. This model achieved a C-statistic of 1681 with 1464 degrees of freedom.
Raising the $R_V$ up to 4 the fit yielded -- instead -- a C-statistic of 1696 for 1464 degrees of freedom. Increasing $R_V$ further to 5.5 (blue line in Fig. \ref{fig:MRNvsHM_WD}) resulted in an even higher C-statistic of 1708. However, models with such high $R_V$ values are rejected by the Akaike test, described in Section \ref{da:dust}.

\paragraph{Hirashita \& Murga.}

Adopting the dust size distribution of \citetalias{hm2020} for a diffuse interstellar medium ($\eta_{\rm{dense}} = 0.1$; purple curve in Fig.~\ref{fig:MRNvsHM_WD}) the fit yields a C-statistic of 1692, while for a dense medium ($\eta_{\rm{dense}} = 0.8$; Fig.~\ref{fig:MRNvsHM_WD}, green curve) the value increases to 1719. Both cases are rejected by the Akaike test. Intermediate fractions provide better fits: for $\eta_{\rm{dense}} = 0.3$ (Fig.~\ref{fig:MRNvsHM_WD}, light purple curve) the C-statistic is 1677, and for $\eta_{\rm{dense}} = 0.5$ (Fig.~\ref{fig:MRNvsHM_WD}, light green curve), which closely resembles the \citetalias{MRN1977} distribution, the C-statistic is 1681. In both Akaike-accepted models, the composition is dominated by amorphous olivine, with amorphous quartz contributing about 20-30\% (Fig.~\ref{risultati_totali}).

\section{Discussion}\label{sec5}

\begin{table*}[h!]
	\centering
    \caption{Abundances and fractional depletions of magnesium and silicon for the good dust size distributions.}
    \begin{tabular}{ccccc} 
 & MRN & WD ($R_V = 3.1$)  & HM ($\eta_{\rm{d}} = 0.3$) & HM ($\eta_{\rm{d}} = 0.5$) \\
\hline
\hline
silicon & & & & \\
\hline
$A_Z/A_{\odot}$ & $1.4\pm0.7$ & $0.7\pm0.4$ & $0.7^{+0.7}_{-0.3}$ & $0.7\pm0.4$ \\
$\delta_Z$ & $0.99\pm0.01$ & $0.97\pm0.01$ & $0.98_{-0.02}^{+0.01}$ & $0.98\pm0.01$\\
\hline
magnesium & & & & \\
\hline
$A_Z/A_{\odot}$ & $0.9\pm0.7$ & $0.9\pm0.5$ & $0.8\pm0.2$ & $0.8\pm0.6$\\
$\delta_Z$  & $0.92^{+0.08}_{-0.09}$ & $0.82\pm{0.10}$ & $0.90\pm0.01$ & $0.88_{-0.17}^{+0.12}$\\
\end{tabular}
\label{table:abundances}
\end{table*}

\subsection{Dust size distribution}

From this analysis, we can rule out the following dust size distribution models along the line of sight of GX 13+1, as they yield $\Delta$AIC values greater than 10 compared to the best-fit model (see Table \ref{table:dustgasNh}): the \citetalias{wd2001} models with $R_V = 4.0$ and $R_V = 5.5$, and the \citetalias{hm2020} models with $\eta_{\rm{dense}} = 0.1$ and $\eta_{\rm{dense}} = 0.8$.
These results allow us to exclude, for this sightline, the presence of a denser-than-average ISM--represented by the \citetalias{wd2001} models with $R_V = 4.0$ and $5.5$, and by the \citetalias{hm2020} model with $\eta_{\rm{dense}} = 0.8$. In the Milky Way, a typical average value is $R_V \simeq 3.1$ \citep{Johnson1968}, which can be used as a reasonable approximation for most lines of sight. This result is consistent with the findings of \citet{Schlafly2016}, who report that fewer than one percent of Galactic sightlines exhibit $R_V > 4.0$. Moreover, these size distributions would imply a substantially higher quartz content in the interstellar medium than typically inferred, raising concerns about their physical plausibility (see e.g. \citealt{Min2007}).

The exclusion of these denser-environment models is also supported by high-resolution X-ray spectroscopy studies in other directions. For example, in their analysis of the Mg and Si K edges toward GX~3+1 \citep{Rogantini2019}, the authors explicitly tested a modified \citetalias{MRN1977} size distribution including an enhanced population of large grains. Although such a model lead to a formally improved fit to the X-ray absorption edges, it required a too high fraction of Mg and Si to remain in the gas phase in order to reproduce the observed spectra.

We can also reject the scenario of an extremely diffuse ISM, represented by the \citetalias{hm2020} model with $\eta_{\rm{dense}} = 0.1$, where grain shattering dominates leading to a dust size distribution favoring smaller grains.

A consistent picture also emerges from X-ray dust scattering halo studies of GX~13+1. In particular, \citet{Clark2018} model the Chandra halo profile of this LMXB using a range of grain size distributions, including the \citet{wd2001} prescription, the BARE-GR-B model of \citet{Zubko2004}, and simple power-law forms. Their best-fitting solutions are obtained when allowing for multi-cloud dust geometries combined with power-law size distributions. Importantly, these models do not require extreme grain growth and remain broadly compatible with standard diffuse-ISM dust properties, reinforcing the conclusion that the dust along this sightline is inconsistent with size distributions typical of either very dense or extremely diffuse environments.

\subsection{Dust composition}

The dust observed along the LOS to GX 13+1 reveals an ISM primarily composed of amorphous olivine, with a significant fraction (20 - 30\%) of fully amorphous quartz. This finding, based on a different dataset, confirms the results of \citet{Zeegers2019} for this direction. Consistent with observations along other sightlines \citep{Zeegers2019, Rogantini2020}, we find no significant contribution from pyroxene; instead, olivine remains the dominant silicate component in the ISM. This dominance of olivine is further supported by infrared studies, such as \citet{Kemper2004}, who reported that approximately 85\% of the silicate dust is in olivine form.

Crystalline or amorphous $\rm{SiO_2}$ (silica) is generally not considered a major constituent of the ISM, but amorphous silica has been sporadically detected. For example, \citet{Fogerty2016} reported up to 20\% silica by mass in warm regions of the Trapezium LOS, based on analysis of the 9.7~$\rm{\mu m}$ feature. Moreover, amorphous silica is also observed in protoplanetary disks \citep{Sargent2009}, suggesting that while it is not a prevalent ISM component, it can appear in specific environments, possibly linked to local dust processing.

\subsection{Depletions and abundances}

From our spectral fits, we derived the depletion fractions of oxygen, magnesium, silicon, and iron - the constituents of our dust models (see Table \ref{table:dustsample}). 
The fractional depletion, is typically expressed as a percentage and is defined as $\delta_{\rm{Z}} = 1 - 10^{D_{\rm{Z}}}$, where $D_{\rm{Z}}$ is determined by comparing the abundance of the gas-phase element Z with its standard solar reference abundance. 

For magnesium and silicon, whose absorption edges we explicitly modeled, we found significant depletion into the solid phase. In our best-fit models, 80-90\% of magnesium and 97-99\% of silicon are locked in dust grains (see Table \ref{table:abundances}).

The depletion fractions of oxygen and iron can only be inferred indirectly from our models and are therefore less tightly constrained. Our results suggest that oxygen is moderately depleted, with less than 30\% incorporated into dust, while more than 80\% of iron appears to be in the solid phase.

The depletion levels derived in this work are consistent with previous X-ray studies (e.g., \citealt{Costantini2012,Pinto2013, Zeegers2019, Rogantini2019, Rogantini2020}) and are also in agreement with the depletions reported by \citet{Jenkins2009} and \citet{Dwek2016}.
Our derived elemental abundances are consistent with those reported in previous studies (e.g., \citealt{Zeegers2019, Rogantini2020}) and do not show significant deviations from solar values (see Table \ref{table:abundances}).
The allowed depletion ranges adopted in these fits are based on \citet{Whittet_book} and \citet{Jenkins2009}, and are summarized in Table \ref{depletions_range}.

\begin{table}[h]
\centering
\caption{Allowed depletion ranges adopted in this work, based on \citet{Whittet_book} and \citet{Jenkins2009}.}
\begin{tabular}{lcc}
\hline
Element & Depletion ranges \\
\hline
Si  & 0.4 -- 1.0 \\
Mg  & 0.4 -- 1.0 \\
O & 0 -- 0.4  \\
Fe & 0 -- 0.2 \\
\hline
\end{tabular}
\label{depletions_range}
\end{table}

\subsection{Dust crystallinity}

The best-fit model suggests a crystalline-to-amorphous dust ratio ($\zeta_1 =$ crystalline dust / (crystalline dust + amorphous dust)) lower than that reported in previous X-ray studies of several LMXBs \citep{Zeegers2019, Rogantini2019, Rogantini2020}, but consistent with values observed in the infrared. In particular, at these wavelengths, the smooth profiles of the features near $\sim 9\,\rm{\mu m}$ and $\sim 18\,\rm{\mu m}$ indicate that less than $2.2\%$ of interstellar dust is crystalline \citep{Kemper2004, Kemper2005}. In this work, we find $\zeta_1 = 0.02$, with an upper bound $\zeta_1 \leq 0.05$.

A key aspect of this work is the significantly improved spectral resolution, made possible by using - for the first time - the third-order spectrum of the \textit{Chandra} HETG. Moreover, such  analysis has been performed on data unaffected by known instrumental features near the Si K edge (see \ref{appA}), ensuring a clearer view of the absorption profile. These improvements enable a more reliable characterization of interstellar dust along this line of sight. 

\section{Conclusions}\label{sec6}

In this work, we used \textit{Chandra} HETG (MEG+1 and MEG$-$3) observations of GX 13+1 to study the composition and size distribution of interstellar dust along its line of sight. By modeling the Si K and Mg K edges with a range of laboratory-based dust species and grain size distributions, we constrained both the physical and chemical properties of the ISM. Our analysis shows that:
\begin{itemize}
\item The classical \citetalias{MRN1977},  the \citetalias{wd2001} ($R_V=3.1$) distributions, along with the \citetalias{hm2020} ($\eta_{\rm{dense}}=0.3-0.5$) models, best reproduce the observed absorption profiles. Distributions representing very dense or highly diffuse ISM are statistically disfavored.
\item The dust on this LOS is predominantly composed of amorphous olivine.
\item The crystalline fraction of dust is low ($\leq 5\%$), bringing the X-ray estimate into agreement, for the first time, with the crystalline fractions previously inferred from infrared observations.
\end{itemize}

These findings confirm the presence of diffuse ISM conditions typical of the Galactic Plane ($R_V \simeq 3.1$) along the sightline to GX 13+1 and emphasize the role of X-ray spectroscopy as a powerful tool for probing the solid phase of the interstellar medium, not only to constrain dust composition, but also to directly test different grain size distributions.
Looking forward, \textit{NewAthena} observations will provide unprecedented improvements thanks to its high effective area. This will enable the detection of Mg and Si absorption features with high signal-to-noise ratios, allowing us to place more stringent constraints on grain composition, crystallinity, and size distributions across different Galactic environments.

\begin{acknowledgements}

We thank Hiroyuki Hirashita for providing the size distribution tables published in Hirashita and Murga. We thank Jon Miller for discussion on the Chandra 3rd order data of this source.
SZ and IAC acknowledge the ASIAA summer student program. SZ acknowledges the ESA Research fellowship. IAC acknowledges support from Fundaci\'on Mauricio y Carlota Botton and the Cambridge International Trust.
This publication was produced while B.V. attending the PhD program in  in Space Science and Technology at the University of Trento, Cycle XXXVIII, with the support of a scholarship financed by the Ministerial Decree no. 351 of 9th April 2022, based on the NRRP - funded by the European Union - NextGenerationEU - Mission 4 "Education and Research", Component 1 "Enhancement of the offer of educational services: from nurseries to universities" - Investment 4.1 "Extension of the number of research doctorates and innovative doctorates for public administration and cultural heritage".
This research has made use of data obtained from the Chandra Data Archive and the Chandra Source Catalog, and software provided by the Chandra X-ray Center (CXC) in the application package CIAO. The Space Research Organization
of the Netherlands is supported financially by NWO.

\end{acknowledgements}

\bibliography{main_AA_formatted}
\bibliographystyle{aa}

\begin{appendix}
\section{Instrumental features near the Si K edge}\label{appA}

In previous studies \citep{Zeegers2019, Rogantini2019, Rogantini2020}, at least an emission instrumental feature near the Si K absorption edge (at $6.742\,\angstrom$) was consistently observed in all sources analyzed, with the notable exception of the +1 order MEG spectra, which did not display any emission in excess. Among the spectral orders used in these studies, the +1 MEG is the only one that does not intersect the front-illuminated CCDs in the silicon region. This instrumental features, detailed in a \textit{Chandra} calibration memorandum\footnote{\url{https://space.mit.edu/CXC/calib/hetg_user.html}}, actually consists of two narrow emission features, one at $6.742\,\angstrom$ and another at $6.714\,\angstrom$. In this calibration memorandum, these lines are believed to originate from fluorescence in the $\rm{SiO_2}$ gate structure present in the front-illuminated CCDs, likely induced by the interaction of X-rays with the detector material.

In our current analysis, we also note that in addition to MEG+1, also for MEG$-$3 the SiK edge energy falls on a back illuminated chip.

The two instrumental features occur near and right in the middle of the edge jump, where the presence of shifts and subtle features can be important in the interpretation of the dust features. Therefore, we only include the MEG+1 and MEG$-$3 in our analysis.

\end{appendix}

\end{document}